\newcount\Comments  % 0 suppresses notes to selves in text
\documentclass[aps,prb, reprint, superscriptaddress]{revtex4-2}
\usepackage[english]{babel}
\usepackage[utf8]{inputenc}
\usepackage{siunitx}
\usepackage[colorinlistoftodos, color=green!40, prependcaption]{todonotes}
\usepackage{graphicx}
\usepackage[colorinlistoftodos]{todonotes}
\usepackage{amsthm}
\usepackage{mathtools}
\usepackage{physics}
\usepackage{xcolor}
\usepackage{graphicx}
\usepackage[left=23mm,right=13mm,top=35mm,columnsep=15pt]{geometry} 
\usepackage{adjustbox}
\usepackage{placeins}
\usepackage[T1]{fontenc}
\usepackage{lipsum}
\usepackage{csquotes}
\usepackage[pdftex, pdftitle={Article}, pdfauthor={Author}]{hyperref} % For hyperlinks in the PDF
\usepackage{xr}
\usepackage{bm}
\usepackage{booktabs}
\makeatletter

\newcommand{\kibitz}[2]{\ifnum\Comments=1\textcolor{#1}{#2}\fi}

\usepackage{natbib}
\begin{document}

\title{Photoinduced phase heterogeneity and charge localization in SnSe}

\author{Benjamin J. Dringoli}
    \affiliation{Department of Physics, McGill University, Montreal, QC, Canada H3A2T8}

\author{Stefano Mocatti}
    \affiliation{Department of Physics, University of Trento, Via Sommarive 14, 38123 Povo, Italy}

\author{Giovanni Marini}
    \affiliation{Department of Physics, University of Trento, Via Sommarive 14, 38123 Povo, Italy}

\author{Zhongzhen Luo}
    \affiliation{Department of Chemistry, Northwestern University, Evanston, IL 60208, United States}
    \affiliation{Key Laboratory of Eco-Materials Advanced Technology, College of Materials Science and Engineering, Fuzhou University, Fuzhou 350108, P. R. China}

\author{Matteo Calandra}
    \affiliation{Department of Physics, University of Trento, Via Sommarive 14, 38123 Povo, Italy}

\author{Mercouri G. Kanatzidis}
    \affiliation{Department of Chemistry, Northwestern University, Evanston, IL 60208, United States}

\author{David G. Cooke}
    \email[Correspondence email address: ] {david.cooke2@mcgill.ca}
    \affiliation{Department of Physics, McGill University, Montreal, QC, Canada H3A2T8}

%\date{\today} % Leave empty to omit a date

\begin{abstract}
\noindent Time-resolved multi-terahertz (THz) spectroscopy is used to observe pump fluence-dependent dynamics in the optical conductivity of photoexcited tin selenide (SnSe) over an ultrabroadband spectral range of 0.5 - 11~THz at fluences from 0.1 - 7.5~mJ/cm$^2$. A free carrier Drude spectrum is observed at pump fluences below 3~mJ/cm$^2$, with optical phonons well described by the equilibrium Pnma structural phase. With increasing fluence, a suppression of the DC photoconductivity is observed, indicating an interruption of long range transport due to phase disorder. Concomitantly, the optical phonons exhibit features that can no longer be explained by a pure Pnma phase, with a frequency shift and narrowing of the $B^2_{1u}$ mode and a new mode appearing at $\sim$3.0~THz consistent with a transition to a higher-symmetry structure. At an intermediate fluence of 3.1~mJ/cm$^2$, a high frequency Lorentzian component consistent with phase heterogeneity appears that rapidly redshifts after 2~ps and whose amplitude exponentially decays on a 90~ps time scale. Our experimental measurements and theoretical calculations provide evidence for a non-thermal, photo-induced nucleation of higher symmetry, semi-metallic phase domains in SnSe appearing within 200~fs.
\end{abstract}

%\keywords{thermoelectric, SnSe, electron-phonon coupling, dynamics, terahertz, spectroscopy, photodoping, phase transition, Pnma, Immm, linewidth, phase transition, symmetry change, phonon degeneracy}

\maketitle

\section{Introduction}

\noindent The fundamental electronic and vibrational properties of tin selenide (SnSe) have recently become a topic of significant interest, after it was shown to be a record-setting thermoelectric material along the in-plane b-axis in its high-temperature $Cmcm$ phase \cite{ZhaoNAT2014,ChangCM2018,AlsalamaRAMS2020,TayariPRB2018}. This performance has been mainly attributed to large acoustic and optical phonon anharmonicity \cite{mingo,PhysRevLett.117.075502,PhysRevLett.122.075901} reducing thermal conductivity and, to a minor extent, to multiple electronic band structure extrema contributing to high electronic conductivity \cite{LiNP2015,HongMTP2019,MaPRB2018}. The origin of this high thermoelectric performance and how it can be enhanced in the room-temperature $Pnma$ phase, shown in Figure \ref{SnSeStruc}(a), has been studied via theoretical modeling \cite{PhysRevLett.122.075901,CarusoPRB2019,SkeltonPRL2016} and structural probes \cite{LiNP2015, BansalPRB2016,HongMTP2019,LangianNC2020, HuangPRL2023}. \\
\indent SnSe is known to exhibit a strong coupling between electronic and lattice structure. Changes to chemical doping \cite{DasJPCM2020,GuoPRB2015,MaedaPRB2018,YanPCCP2016,ZhaoSCI2016} or pressure \cite{EfthimiopoulosPCCP2019,BiesnerNPGAM2021,GusmaoCMS2018,NishimuraPRL2019,PalPRB2020} were shown to modify Fermi surface topology in the form of Lifshitz transitions and lattice properties such as phonon-phonon scattering. Under non-equilibrium conditions following optical excitation, evidence for an ultrafast transition to the high temperature $Cmcm$ phase was reported at lower excitation fluences of $\sim$20~$\mu$J/cm$^2$ \cite{HanJPCL2022}. At higher excitation fluences on the order of mJ/cm$^2$, an ultrafast transition towards a uniquely photoaccessible $Immm$ phase was proposed \cite{HuangPRX2022}. The transient $Immm$ phase was shown to have reduced phonon anharmonicity measured via time-resolved X-ray diffraction \cite{HuangPRL2023}. Studies also reported domain nucleation \cite{WangNPJQM2021} and evidence for band gap collapse to a semi-metallic phase at a critical fluence on the order of several mJ/cm$^2$ \cite{DringoliPRL2024,MogiARX2025,MocattiJPCL2023}. Importantly, a photoinduced transition towards a higher-symmetry $Immm$ phase may hint at a potential route to a meta-stable or even fully stable photo-induced topological crystalline insulator in the $Fm\bar{3}m$ rocksalt phase \cite{MocattiJPCL2023}. To date there has been no evidence of a complete $Pnma \rightarrow Immm$ phase transition. Recent time-resolved angle-resolved photoemission spectroscopy (TR-ARPES) measurements indirectly support a nucleation of local $Immm$ domains with a $Pnma$ background through the observation of photoexcited Dirac-like states \cite{MogiARX2025}. Such phase boundaries are expected to affect long-range carrier transport properties that could be identified through THz measurements. 
\\
\indent In this work, we apply time-resolved multi-THz spectroscopy, shown in Fig.~\ref{SnSeStruc}(b), to simultaneously probe the photoinduced dynamics of long range free carrier transport, optical phonons, and collective plasmonic resonances arising from phase heterogeneity over more than a decade of energy (2 - 45 meV) with $\sim$40~fs temporal resolution \cite{JepsenLPR2011}.
Furthermore, we perform electronic structure calculations to interpret the experimental results.

The paper is organized as follows. In Sec.\ref{sec:Technical_details}, we provide details of the experimental measurements and theoretical calculations. In Sec.\ref{sec:Results}, we present the experimental results and compare them with density-functional calculations. In Sec.\ref{sec:Discussion}, we discuss the implications of the measured and computed quantities. Finally, in Sec.\ref{sec:Conclusions}, we summarize the main conclusions of the present work.

\section{Technical details}
\label{sec:Technical_details}

\subsection{Experiments}

The ultrafast THz response after photoexcitation with 35~fs-duration, 800~nm optical pump pulses was measured via air-plasma THz generation and detection \cite{KimOE2007,DaiPRL2006,DAngeloOSA2016,HoOE2012}, allowing for probe bandwidths spanning 0.5 - 11~THz to be captured with sub-picosecond pump-probe resolution. The SnSe sample was a $2.5$~mm $\times$ $2.5$~mm $\times$  $500$~nm exfoliated single crystal of SnSe grown using methods described in Ref.~\cite{ZhaoNAT2014}, and from the same parent crystal as in Refs.~\cite{ZhaoNAT2014,ReneDeCotretPNAS2022,DringoliPRL2024}. The thickness of the exfoliated SnSe was measured via atomic force microscopy, with measurements shown in Supplemental Figure 6. The exfoliated crystal was mounted on a 500 $\mu$m-thick electronic-grade diamond substrate, used for its constant THz refractive index, low optical absorption, high thermal conductivity, and negligible photoconductivity from shallow boron impurities (here ppb).

\begin{figure}[t!]
    \centering
    \includegraphics[width=\columnwidth,trim={7cm 0cm 7cm 0cm},clip]{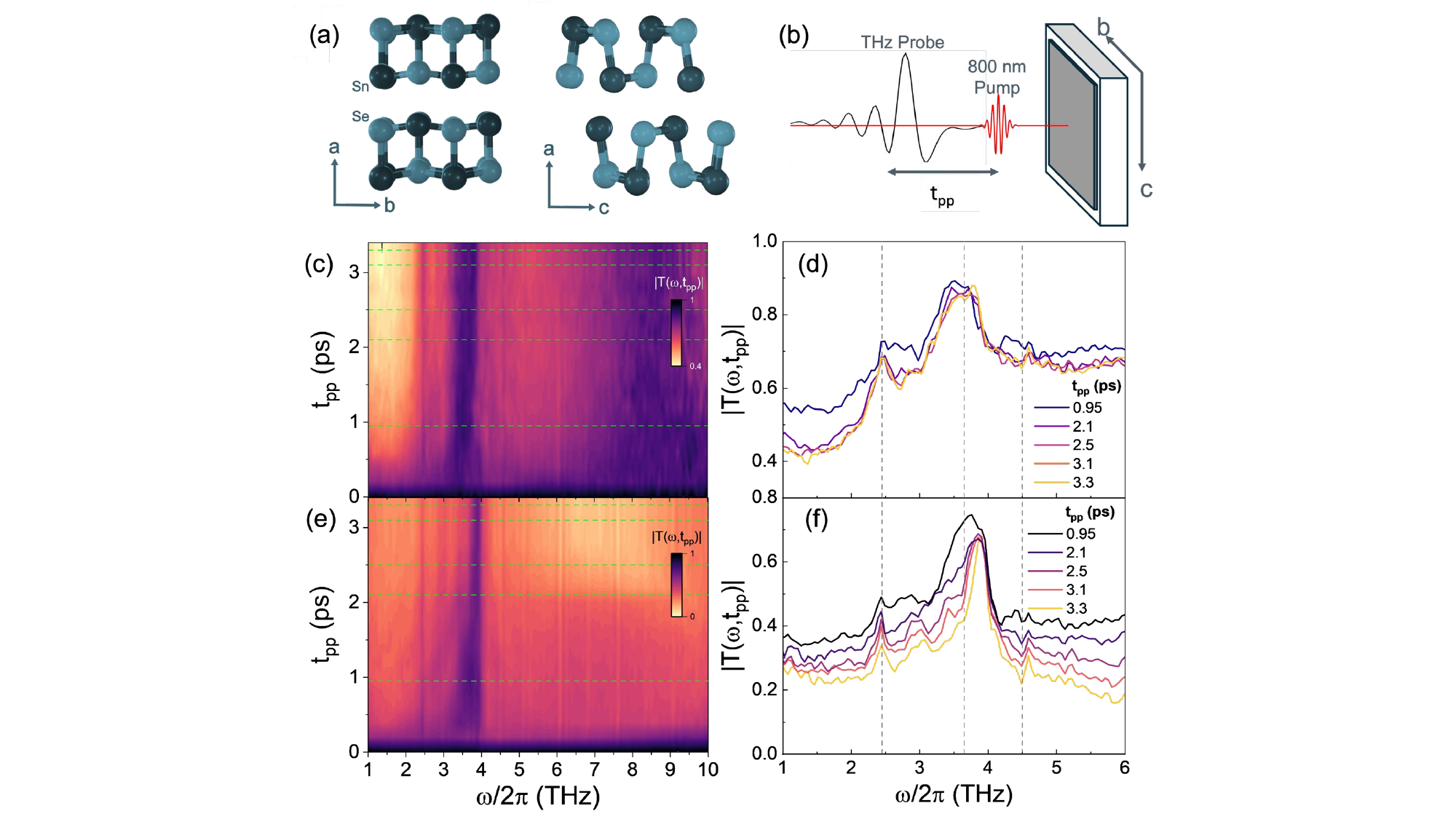}
    \caption{(a) SnSe $Pnma$ atomic arrangements showing the two in-plane crystal axes, zig-zag (b-axis) and armchair (c-axis). (b) Experiment schematic showing the optical pump and THz probe polarization parallel to the c-axis of the exfoliated SnSe mounted on diamond. t$_{pp}$ is the pump-probe delay. Two-dimensional maps of the amplitude of the THz transmission coefficient $|\tilde{T}(\omega,t_{pp})|$ are shown for pump fluences of (c) 1.3 and (e) 3.1~mJ/cm$^2$. Constant $t_{pp}$ cuts in are shown in (d) and (f) for the maps in (c) and (e), respectively.}
    \label{SnSeStruc}
\end{figure}

\begin{figure*}[th!]
    \centering
    \includegraphics[width=2\columnwidth]{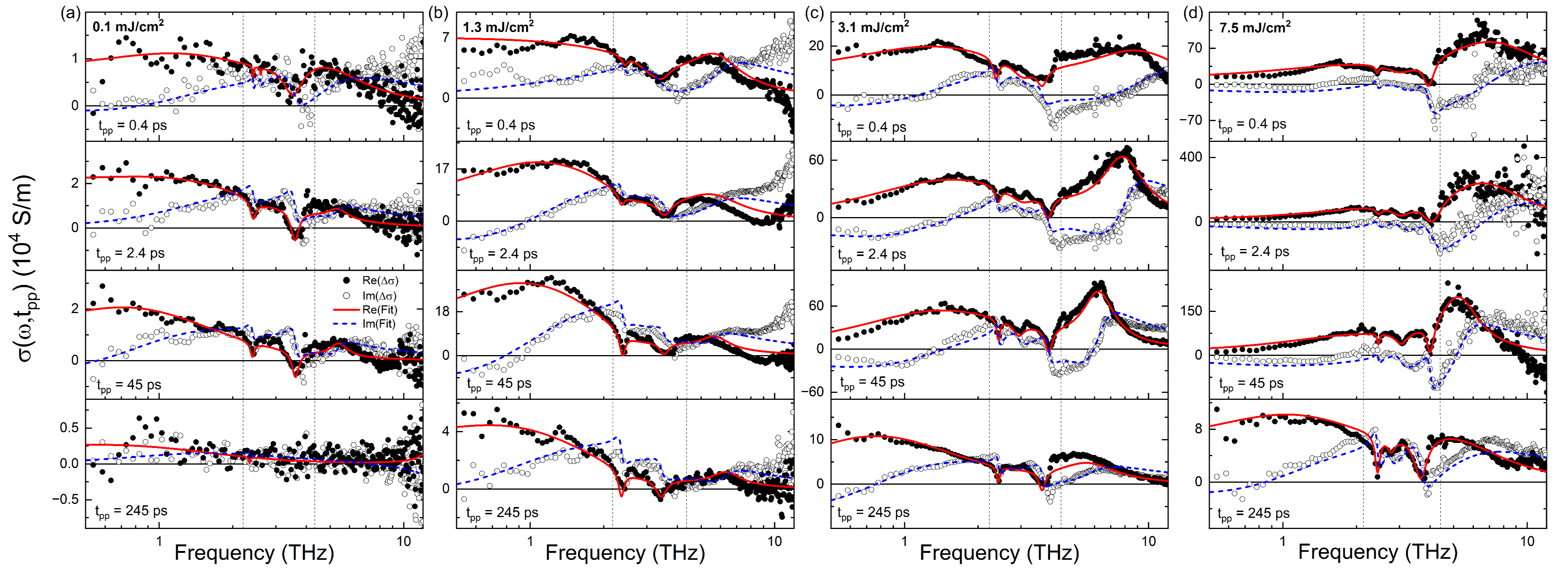}
    \caption{Real (solid symbols) and imaginary (hollow symbols) components of $\tilde{\sigma}(\omega,t_{pp})$ with full conductivity spectra fits for multiple time delays after (a) 0.1, (b) 1.3, (c) 3.1, and (d) 7.5~mJ/cm$^2$ pump excitation. The three fit regions are delineated by dashed lines and sequentially fit with Drude-Smith, three Lorentzian, and single Lorentzian models, respectively. Blueshifting of the low-frequency conductivity peak and phonon feature narrowing is seen with increasing fluence, which returns for $t_{pp} >$ 200~ps. At higher fluences a large peaked enhancement is seen above all phonon frequencies which exhibits strong center frequency change with both fluence and pump-probe delay.}
    \label{AllSpectra}
\end{figure*}

%\section{Methods}

\indent The SnSe-on-diamond was placed in a sample-in-vacuum cryostat (Janis ST-300MS) for measurements at 80~K to better isolate the pump-induced non-thermal response. The sample was measured at normal incidence with the THz and pump pulses linearly polarized along the SnSe c-axis, where the lattice modes exhibit the strongest anharmonicity that contributes to the thermal behavior near the $Pnma \rightarrow Cmcm$ phase transition \cite{ChattopadhyayJPCS1986}. The transmitted THz electric field was sampled in time $t$ using a double-modulation scheme to extract the transmitted THz field with optical excitation $E_p(t,t_{pp})$ and the optical pump-induced differential between pumped and reference fields $\Delta E(t,t_{pp}) = E_p(t,t_{pp}) - E_r(t)$ at various pump probe times $t_{pp}$.

The complex-valued THz transmission coefficient given by
\begin{equation}
\begin{split}
    \tilde{T}(\omega,t_{pp}) = \frac{ |E_p(\omega,t_{pp})|}{|E_r(\omega)|} \big[ \cos\phi(\omega,t_{pp})+ i \sin\phi(\omega,t_{pp}) \big]
\end{split}
\end{equation}
where $\phi = \phi_p - \phi_r$, can then be extracted from the measured 2D time-domain THz fields using Fourier analysis. Figure~\ref{SnSeStruc}(c) shows an example $|\tilde{T}(\omega,t_{pp})|$ map at a pump fluence of 1.3~mJ/cm$^2$ with cuts shown in Fig.~\ref{SnSeStruc}(d), exhibiting IR-active c-axis polarized phonon modes: $\text{B}^1_{1u}$ at $\omega_{0,1} =$ 2.4~THz 
and $\text{B}^2_{1u}$ at $\omega_{0,2} =$ 3.7~THz \cite{EfthimiopoulosPCCP2019}. At a higher fluence of 3.1~mJ/cm$^2$ shown in Fig.~\ref{SnSeStruc}(e,f), the $\omega_{0,2}=$ 3.7~THz B$^2_{1u}$ feature shows dynamic, pump-induced changes to both linewidth and center frequency indicating strong electron-phonon coupling \cite{CarusoPRB2019}.  It is clear then that the lattice response cannot be treated solely as a static background but evolves dynamically under photoexcitation.

$\tilde{T}(\omega,t_{pp})$ can then be further transformed into the time-dependent optical conductivity $\tilde{\sigma}(\omega,t_{pp}) = \sigma_1(\omega,t_{pp})+i\sigma_2(\omega,t_{pp})$ as \cite{HegmannSPIE2002} 

\begin{equation}
   \tilde{\sigma}(\omega,t_{pp}) = \frac{n + 1}{Z_0 d} \left( \frac{1}{\tilde{T}(\omega,t_{pp})}-1 \right),
\end{equation}

\noindent where $d = 60$~nm is the pump penetration depth at $\lambda = 800$~nm for SnSe \cite{MakinistianPSSB2009}, $Z_0 = 377~\Omega$ is the impedance of free space, and $n = 2.38$ is the THz refractive index of the diamond substrate. Figure~\ref{AllSpectra} shows the conductivity spectra for all pump fluences investigated at several pump-probe delays, along with fit curves described below.\\
\begin{table*}[th!] % [H] float option can help with specific placement but use sparingly
    \centering
    \caption{Explanation of frequency ranges and descriptions of Lorentzian fit components. $L=$~1 - 3 are used for phonon resonances between 2 - 4.4~THz, $L=$~4 is used for a high-frequency plasmonic-type response. Each feature can be observed in the full-spectrum fits presented in Figure \ref{AllSpectra}.}
\begin{center}
\begin{tabular}{lcccc}
 \toprule % Top horizontal line
\textbf{L} & \bm{$\omega_{p,L}$} \textbf{(THz)} & \bm{$\omega_{0,L}$} \textbf{Range (THz)} & \bm{$\gamma_L$} \textbf{(ps$^{-1}$)} & \textbf{Description} \\ 
 \midrule % Line separating headers from data
1 & $\omega_{p,1}$ & 2.2 $<\omega_{0,1}<$ 2.6 & $\gamma_1$ & $B^1_{1u}$ Phonon \\ 
2 & $\omega_{p,2}$ & 3.4 $<\omega_{0,2}<$ 4.5 & $\gamma_2$ & $B^2_{1u}$ Phonon \\ 
3 & $\omega_{p,3}$ & 2.7 $<\omega_{0,3}<$ 3.3 & $\gamma_3$ & New Phonon Feature \\ 
4 & $\omega_{p,4}$ & 4.5 $<\omega_{0,4}<$ 11 & $\gamma_4$ & Plasmonic Peak \\ 
\end{tabular}
\end{center}
\end{table*}

\indent To analyze the recorded data, the recovered $\tilde{\sigma}(\omega,t_{pp})$ spectra are divided into 3 main regions: long-range transport (0.5 - 2~THz), lattice modes (2 - 4.5~THz), and high-frequency electronic response (4.5 - 11~THz), shown by dashed vertical lines in Figure \ref{AllSpectra}. In the low-frequency region, a Drude-Smith model \cite{SmithPRB2001} was used to track long-range carrier conductivity with the potential for carrier localization:
\begin{equation}
    \tilde{\sigma}(\omega) = \epsilon_0\epsilon_\infty \omega_{p}^2 \tau \left( \frac{1}{1-i \omega \tau} + \frac{c}{(1-i \omega \tau)^2} \right).
    \label{DSModel}
\end{equation}
\noindent Here $\omega_p$ is the plasma frequency $\omega_p = \sqrt{n e^2/\epsilon_{\infty}\epsilon_0 m^*}$, $\epsilon_{\infty} = $16.2 is the high-frequency relative permittivity \cite{EfthimiopoulosPCCP2019}, $\tau$ is the carrier momentum scattering time, and $c$ is a dimensionless parameter describing the backscattering influence. We are unable to distinguish between electrons and holes in the conductivity response as they both contribute. The advantage of the simple Drude-Smith model is that it can describe continuous transitions across an insulator to metal transition while obeying Kramers-Kronig relations over the entire range and with minimal fit parameters \cite{CookePRB2006}. This includes carrier backscattering from microscopic domains comparable to the carrier mean free path that can suppress low-frequency (long-range) charge transport \cite{CookePRB2006,CockerPRB2017}, making it suitable to describe carrier percolation in disordered media \cite{WaltherPRB2007}.\\
\indent Similarly, a series of Lorentzian models were used to fit individual phonon features ($L=$ 1 - 3) as well as a high-frequency resonance observed in a previous work ($L=$ 4) \cite{DringoliPRL2024}
\vspace{-8pt}
\begin{equation}
    \sigma_L(\omega) = \sum_{L=1}^4 \frac{ \epsilon_0 \omega_{p,L}^2 \omega}{i(\omega_{0,L}^2 - \omega^2)+\omega \gamma_L},
\end{equation}
\noindent with Lorentzian plasma frequencies $\omega_{p,L}$, center frequencies $\omega_{0,L}$, and linewidths $\gamma_L$. Descriptions for each Lorentzian component are given in Table I. The combination of these models allowed for the full complex conductivity spectrum to be globally fit over more than a decade of probe energy with a minimal number of fit parameters.

\subsection{Theory}

The calculations are carried out within density-functional theory (DFT) using \textsc{Quantum ESPRESSO} \cite{Giannozzi2009,Giannozzi2017}. Optimized norm-conserving Vanderbilt pseudopotentials \cite{Hamann2013}, including semicore states, are employed with a kinetic-energy cutoff of 80 Ry for structural relaxations and stress calculations. Exchange and correlation are treated within the generalized-gradient approximation in the Perdew–Burke–Ernzerhof (PBE) form \cite{Perdew_1996}. Since interlayer interactions play an important role in SnSe, van der Waals corrections are included using the Grimme-D3 scheme \cite{Grimme2010}. Brillouin-zone integrals are evaluated on Monkhorst–Pack grids \cite{Monkhorst1976}. A $4\times 12\times 12$ mesh is used for total-energy calculations and structural relaxations, providing a total-energy accuracy better than 1 meV/atom. For phonon calculations in the presence of photocarriers, denser $7\times 18\times 18$ meshes are employed. The effect of spin-orbit coupling on the structural properties and phonon frequencies of the $Immm$ and $Fm\bar{3}m$ phases is found to be small and is therefore neglected in the vibrational calculations.\\
\indent Photoexcited carriers are described within constrained density-functional theory (cDFT) and constrained density-functional perturbation theory (cDFPT) \cite{Marini2021}. The excited state is modeled as a thermalized electron–hole plasma, represented by two Fermi–Dirac distributions with distinct chemical potentials. The photocarrier concentration is denoted by $n_e = x$ e$^-$/u.c., where x is the number of promoted electrons per unit cell. In the cDFT/cDFPT calculations, Marzari–Vanderbilt smearing \cite{MarzariVanderbilt1999} with $\sigma_1 = \sigma_2 = 0.01$ Ry is used in both structural and phonon calculations.\\
\indent Phonon spectral functions are computed including anharmonic effects within the stochastic self-consistent harmonic approximation (SSCHA) \cite{Monacelli_2021}. This framework is combined here with cDFT and cDFPT to account for light-induced anharmonicity. SSCHA calculations are performed on a $2\times 3\times 3$ supercell containing 144 atoms. The spectral functions are computed within the dynamic bubble approximation. Modifications to peak intensities arising from static and dynamic Born effective charges are not included; only infrared-active (IR) B$_{1u}$ modes are considered, consistent with the experimental selection rules adopted here.\\
\section{Results}
\label{sec:Results}
\noindent Selected pump-probe delays with full-spectrum fits of $\tilde{\sigma}(\omega,t_{pp})$ for all excitation fluences are shown in Figure \ref{AllSpectra}. In photoexcited semiconductors, the THz conductivity is typically dominated by a free carrier Drude response corresponding to $c=0$ in the Drude-Smith model in Eqn.~\ref{DSModel} \cite{SmithPRB2001}, with $\sigma_1$ peaked at $\omega=0$. Instead, measurements with increasing pump fluences show a suppression of the low frequency photoconductivity and a blueshift of the conductivity peak away from $\omega=0$. Such a suppression and shift is commonly observed in nanoscale disordered media where domain boundaries act as scatterers in mixtures of insulating and metallic domains, and is particularly sensitive to a percolative transition as conducting domains form a connected pathway over ac conductivity length scales given by $L_\omega=\sqrt{D/\omega}$ where $D$ is the diffusion coefficient and $\omega$ is the THz frequency. The observation of such a suppression indicates the presence of light-induced disorder in photocarrier conduction, recently observed in photoexcited type II Weyl semi-metal WTe$_2$ \cite{ZhouPRB2025}.

\begin{figure}[t!]
    \centering
    \includegraphics[width=0.8\columnwidth,trim={8cm 11.5cm 2cm 2.5cm},clip]{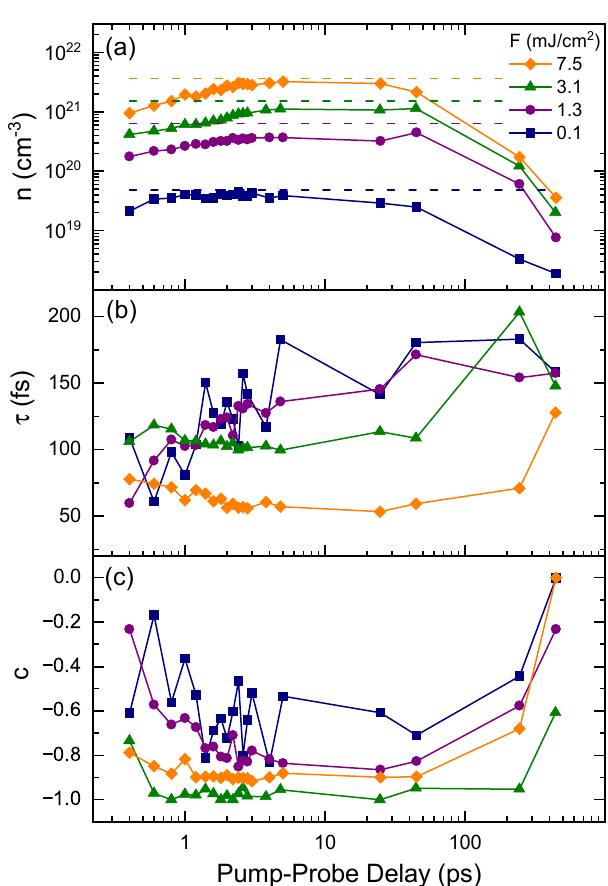}
    \caption{(a) Carrier density calculated from the fitted Drude-Smith plasma frequency ($\omega_{p}$) dynamics. Maximum values agree with those calculated from the input fluences, shown in dashed lines. (b) Drude-Smith scattering rate ($\tau$) showing increased scattering and slower dynamics with increasing fluence. (c) Backscatter parameter $c$ across measured fluences.}
    \label{DSparams}
\end{figure}
\indent We quantify these effects through Drude-Smith fit parameters shown in Figure~\ref{DSparams}. For all fluences, the deposited carrier density $n$ calculated from the plasma frequency $\omega_{p}$ using an effective mass of $m_{e,h}^* =$ 0.2$m_e$ \cite{GuoPRB2015,ShiJAP2015} shown in Figure \ref{DSparams}(a) reaches a maximum 2 - 4~ps following excitation, shifting to longer times at higher fluences $F$. This maximum fitted density aligns with deposited carrier density estimations based on the input fluence, calculated as $n=F\lambda (1-R)/h c d$ and shown in Fig.~\ref{DSparams}(a) as dashed lines. The rise time scale is much longer than the 40~fs temporal resolution of the TRTS system and indicates long-range conductivity dynamics are dominated by slower processes such as inter-valley scattering or charge transfer between domains. This is in agreement with our previous work investigating conductivity dynamics at frequencies above the optical phonons, where inter-valley scattering processes from the SnSe $\Gamma$ band with a larger density of states to surrounding band extrema were posited to explain ps-scale dynamics in the optical conductivity \cite{DringoliPRL2024}.
\\
\indent The fluence-dependent dynamics of the momentum scattering time $\tau$ are shown in Figure \ref{DSparams}(b), exhibit a general reduction with increasing fluence which could be attributed simply to increased electron-hole scattering at such high densities or nanoscale domain formation. Using an effective mass $m^*_{e,h}=$ 0.2$m_e$ for both electrons and holes \cite{GuoPRB2015,ShiJAP2015} gives a microscopic mobility ranging from 450 - 1600~cm$^2$/Vs. This THz ac mobility can be viewed as intrinsic to SnSe within the microscopic grains of the crystal, and so is expected to be higher than reported dc mobilities of 250~cm$^2$/Vs \cite{ZhaoEES2016}. As charge carriers relax with increasing pump-probe delay time, $\tau$ increases for all fluences, with the increase setting in first for lower initial densities at 0.1~mJ/cm$^2$ and 1.3~mJ/cm$^2$ but showing extended suppression at the higher fluences. This indicates electron-hole scattering is the main mechanism for momentum relaxation.
\\
\indent A significant reduction in long range transport, parameterized by $c\rightarrow -1$, is observed for $t_{pp}<5$~ps after photoexcitation for all excitation fluences, shown in Fig.~\ref{DSparams}(c). The general trend of $c\rightarrow -1$ with increasing pump fluence points to this backscattering effect intensifying as more charge carriers are excited. This trend is not explicitly followed at the highest excitation fluence of 7.5~mJ/cm$^2$, which has a slightly smaller $c$ across all pump-probe delays with the current fit model. This may imply a transition to a different conduction mechanism occurring between 3.1 and 7.5~mJ/cm$^2$, perhaps as the system undergoes a percolative transition. This transition to disordered conduction occurs fastest for $F=3.1$~mJ/cm$^2$, within $t_{pp}<0.5$~ps. This localization threshold was confirmed through repeated spectroscopy at fixed pump-probe delays with increasing fluence, shown in Supplemental Figure 3.
\\
\indent Fluence- and time-dependent conductivity dynamics are also present in the region of the spectrum dominated by IR-active optical phonons between 2 - 4.5~THz. The strongest negative peaks correspond to the two known IR-active c-axis phonon modes mentioned previously ($L=$ 1,2). A third B$^3_{1u}$ mode at approximately 4.5~THz is not resolved due to its much lower Born effective charge \cite{EfthimiopoulosPCCP2019}. These peaks are negative due to the neglect of the background phonon conductivity in the thin film reference. However, due to the dynamics present in these features and the sensitivity of the extracted conductivity to the background dispersion, we opt to include localized Lorentzians in our global fit to parametrize the observed changes. Further discussion of this effect can be found in the Supplemental Information.

\begin{figure}[t!]
    \centering
    \includegraphics[width=0.8\columnwidth,trim={0cm 0.5cm 0cm 2cm},clip]{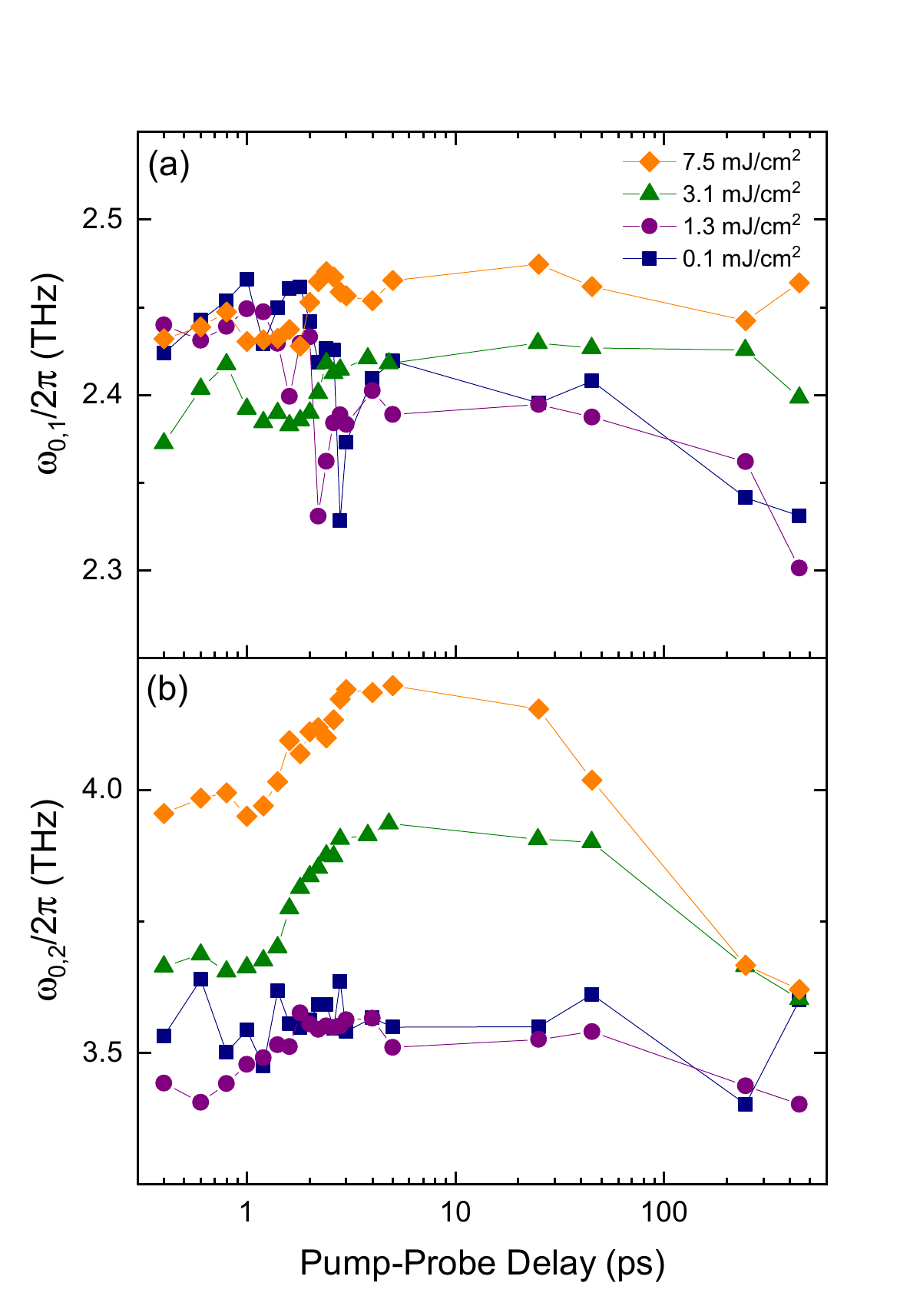}
    \caption{Fitted $\omega_0$ dynamics for the (a) 2.4~THz B$^1_{1u}$ and (b) B$^2_{1u}$ 3.7~THz modes. Strong blueshifting is seen with increasing fluence for the B$^2_{1u}$-centered feature, but only minor blueshifting is seen for the B$^1_{1u}$ feature. The blueshifting follows similar dynamics to the other fitted regions, supporting strong interactions between lattice modes and excited carriers.}
    \label{PhononFits}
\end{figure}

\indent For photoexcitation at lower pump fluences, the B$^2_{1u}$ phonon feature remains quite broad for 10's of ps post-excitation such that the Lorentzian fit cannot accurately capture the linewidth, as shown in Supplemental Figure 4. At higher fluences, however, the phonon feature is seen to narrow considerably upon photoexcitation, which can be directly observed from the transmission map in Figure \ref{SnSeStruc}(e). This dynamic narrowing with increasing fluence is counter to the general expectation that phonon features broaden as excited photocarriers transfer their energy to the lattice \cite{TangPRB2010}. While the B$^2_{1u}$ mode dynamically narrows, the $\omega_{0,1}=$ 2.4~THz B$^1_{1u}$ mode exhibits negligible linewidth or frequency shift, shown in Figure \ref{PhononFits}(a). This narrowing relaxes for $t_{pp}>$ 250~ps, shown in the bottom panels of Figure 2 and pointing to these photoinduced changes being reversible even at the highest fluence tested here. 

\indent The fitted $\omega_{0,2}$ of the B$^2_{1u}$ feature, shown in Figure \ref{PhononFits}(b), also demonstrates dynamics. At $F\leq1.3$~mJ/cm$^2$ fluences the center frequency is largely stable near 3.5~THz, whereas for higher fluences there is a overall and dynamic blueshift on time scales matching the rise in free carrier concentration $n$ shown in Figure~\ref{DSparams}(a). Interestingly, this blueshift is not fully complete before $t_{pp}=$ 1~ps like in the case of the Drude-Smith low-frequency suppression at these fluences, but reaches a maximum 4 - 5~ps after photoexcitation. This blueshift is markedly stronger in the case of 7.5~mJ/cm$^2$ pumping, linking this process directly to the density of photoexcited carriers.

\indent Of note is the presence of an additional feature in the phonon frequency region ($L=$ 3), occurring near 3.0~THz in the 3.1 and 7.5~mJ/cm$^2$ datasets. While at early times the feature is influenced by systematic time-frequency artifacts that are the subject of another publication, this feature is seen to last for 10's to 100's of ps, as shown in Figures \ref{AllSpectra}(c,d). This 3.0~THz mode does not correspond to any identified IR-active mode in either the $Pnma$ or $Cmcm$ phase of SnSe \cite{EfthimiopoulosPCCP2019,LuPRB2019}, and so may be another signature of lattice symmetry change during the relaxation of photoexcited carriers.

\begin{figure*}[t!]
    \centering
    \includegraphics[width=\textwidth,trim={0cm 0cm 0cm 0 cm},clip]{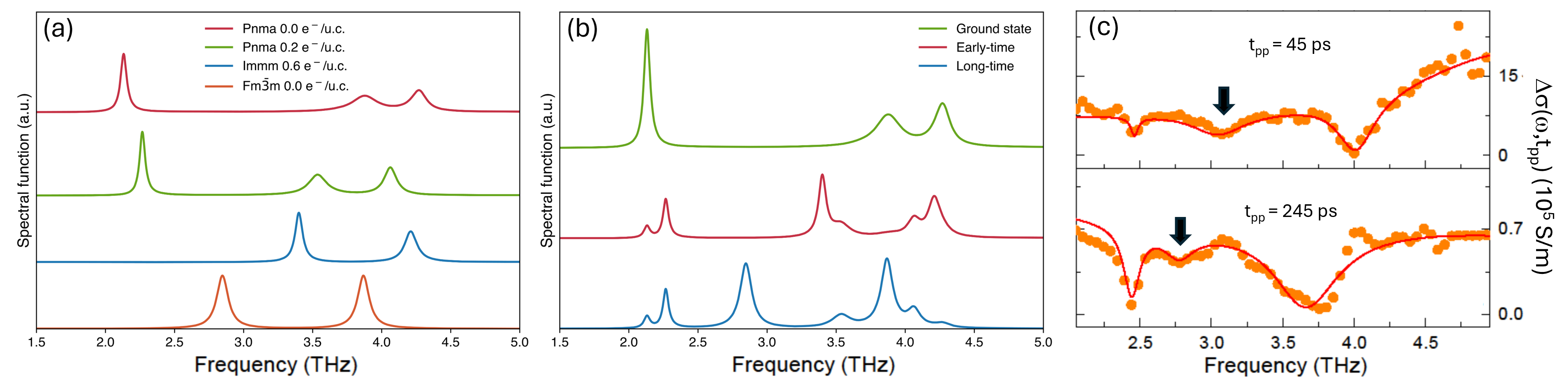}
    
    \caption{(a) Spectral functions for $Pnma$ (ground state and photoexcited), $Immm$ (photoexcited), $Fm\bar{3}m$ (photoexcited) restricted to the IR active B$_{1u}$ modes. (b) Spectral function restricted to the IR active B$_{1u}$ modes of the $Pnma$ (ground state) and after photoexcitation at short and long times. These plots assume a $Pnma \rightarrow Immm \rightarrow Fm\bar{3}m$ partial phase transition pathway. Note the additional features in the 3.0-3.5~THz region where the optical conductivity is measured. (c) Experimental $\sigma_1(\omega,t_{pp})$ at a fluence of 7.5~mJ/cm$^2$ showing the similar development and red-shift of a new Lorentzian dip, assigned to the new $B_{1u}$ mode appearing in (b).}
    \label{SFig5}
\end{figure*}

To clarify this point, we compute the spectral functions of SnSe restricted to the IR-active B$_{1u}$ modes, shown in Fig.~\ref{SFig5}(a) for the ground-state $Pnma$ phase, photoexcited $Pnma$ at 0.2 e$^-$/u.c., photoexcited $Immm$ at 0.6 e$^-$/u.c., and the unexcited $Fm\bar{3}m$ phase. All peaks in the spectral function are infrared active, although their relative intensities depend on the static (ground-state) and dynamical (excited-state) Born effective charges which are neglected in the present treatment. As a consequence, the relative intensities of the theoretical and experimental peaks differ quantitatively; for example, the B$_{1u}^3$ phonon at $\sim$ 4.3 THz is too weak to be observed in the THz spectra.

The effect of photodoping in the $Pnma$ phase is to harden the sharp $B^1_{1u}$ phonon slightly while softening the $B^2_{1u}$ phonon. Photoexcitation at fluences larger than the critical fluence of $0.6$~e$^-$/u.c. creates a new mode between the $Pnma$ $B^1_{1u}$ and $B^2_{1u}$ modes at $\sim3.4$~THz which we assign to the new transient phonon observed in the experimental data above $F_c\approx3$~mJ/cm$^2$ and consistent with a heterogeneous mix of $Pnma$ and $Immm$ phases. 

To simulate the time evolution of the phonon spectra through a partial phase transition, we model in Fig.~\ref{SFig5}(b) the spectral function of a mixed phase. Specifically, we assume that the ground-state spectral function (green) consists exclusively of unexcited $Pnma$, whereas the early-time spectrum (red) is composed of 10\% unexcited $Pnma$, 30\% excited $Pnma$, and 60\% excited $Immm$. The long-time spectrum (blue), instead, is modeled as 10\% unexcited $Pnma$, 30\% excited $Pnma$, and 60\% unexcited $Fm\bar{3}m$. This construction is intended to reproduce qualitatively the photoinduced pathway $Pnma \to Immm \to Fm\bar{3}m$; see Ref.\cite{MocattiJPCL2023} for further details.

For early times following photo-doping, a new mode at $\sim3.4$~THz appears which we identify with the new mode observed in our experiments, shown in the $\sigma_1$ data in Fig.~\ref{SFig5}(c) at $F=7.5$~mJ/cm$^2$. The long-time evolution following relaxation of charge carriers is shown for $t_{pp}=45$ and 245~ps after excitation, showing the red-shift as carriers relax. Here the observed dip in the conductivity could signal the partial formation of the meta-stable $Fm\bar{3}m$ phase. Such a meta-stable phase would appear in the background conductivity discussed previously and would cause a new dip in the conductivity spectrum.

\begin{figure}[th!]
    \centering
    \includegraphics[width=0.8\columnwidth,trim={0.1cm 0.5cm 0cm 2.75cm},clip]{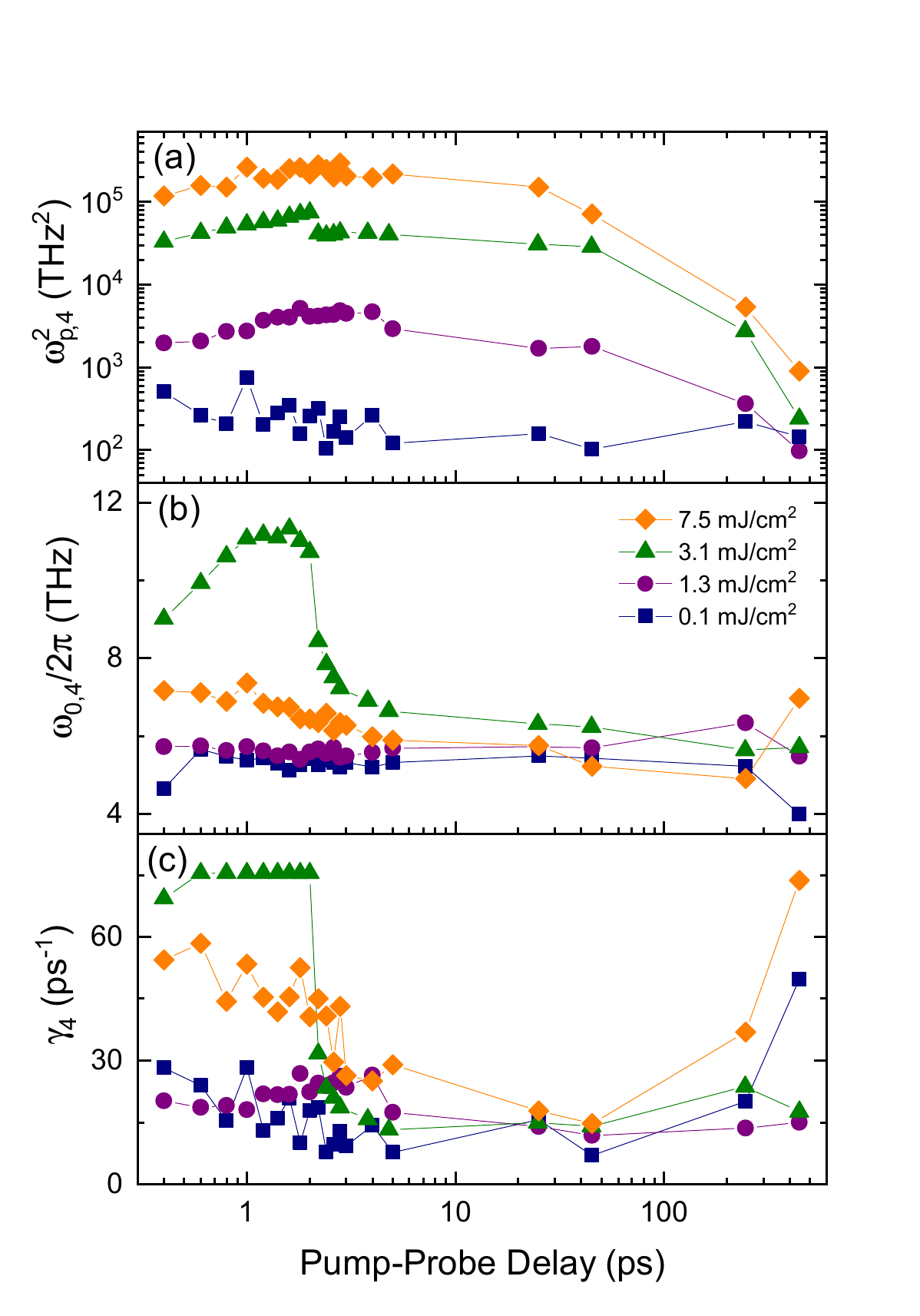}
    \caption{Plasmonic region fitted Lorentzian (a) plasma frequency $\omega_{p,4}$, (b) center frequency $\omega_{0,4}$, and (c) linewidth $\gamma_4$ dynamics across measured fluences. The plasma frequency scaling shows that this high-frequency enhancement scales with input fluence. The center frequency dynamics shows minor changes at the lowest fluences but is most pronounced for 3.1~mJ/cm$^2$ pumping. The linewidth follows similar dynamics to the center frequency, but required a strict upper bound for 3.1~mJ/cm$^2$ due to the presence of a `shelf-like' feature for $0 < t_{pp} < 2$~ps.}
    \label{LorentzFitParams}
\end{figure}

\indent In the high-frequency section of the spectrum between 4.5 - 11~THz, a distinctly non-Drude peak in the optical conductivity appears. This is most clearly visible for 3.1~mJ/cm$^2$ and 7.5~mJ/cm$^2$ pump fluences. Similar to our previous reflection measurements on bulk single crystals\cite{DringoliPRL2024}, this plasmonic-type feature redshifts with increasing pump-probe delay. In these new thin film measurements that are less influenced by the phonon dispersion, we can now see that the resonance appears to settle just above the highest optically active phonon energy. This feature is long-lived, being present for 100's of ps after photoexcitation despite the recovery of zero-frequency conductivity between $t_{pp} =$ 45~ps and $t_{pp}=$ 245~ps. These dynamics are quantified by Lorentzian fit parameters ($L=$ 4), shown in Figure \ref{LorentzFitParams}. Like the pump-induced change to the B$^2_{1u}$ resonance, the high-frequency enhancement is stronger for higher fluence pumping above 3~mJ/cm$^2$, but here the 3.1~mJ/cm$^2$ case shows the largest change in the fitted center frequency $\omega_{0,4}$. In the case of the lower fluences, this response is minor and stays pinned just above the highest active phonon resonance, around 5~THz. 

\indent Additionally, the 3.1~mJ/cm$^2$ high-frequency optical conductivity deviates from a peaked Lorentzian shape at early times, shown most clearly in the $t_{pp} =$ 0.4~ps panel in Figure \ref{AllSpectra}. This initial `shelf'-like conductivity only appears as a deviation from the fit for the first $\sim$2~ps post-excitation before being superseded by the large redshifting Lorentzian peak, but the presence of such a feature may explain the steep slope on the high-frequency edge of the B$^2_{1u}$ phonon at later times when the high-frequency enhancement has largely relaxed.

\section{Discussion}
\label{sec:Discussion}

\noindent The observed changes in each region support the formation of localized photoexcited domains among a $Pnma$ background. The continued observation and lack of strong dynamics of the equilibrium 2.4~THz B$^1_{1u}$ phonon mode is a clear signature that the $Pnma$ phase is still present, and the suppressed low-frequency conductivity well-described by the Drude-Smith model points to reduced mean free paths after photoexcitation \cite{JepsenLPR2011}. This suppression of low-frequency conductivity, which is dominated by long-range transport, suggests domain nucleation following photoexcitation. Further photoexcitation increases the volume fraction of these domains, resulting in increased charge carrier localization parametrized through the Drude-Smith $c$. It is possible that the reduction of the $c$ parameter magnitude at the highest fluence represents a crossover where photoinduced domains become dense enough to re-allow long-range transport through a percolative transition of the photoexcited phase. The recovery of pure Drude conductivity with a maximum near $\omega=0$ at long pump-probe delays indicates a continuous relaxation back to a homogeneous $Pnma$ semiconducting phase over time. These results corroborate recent TR-ARPES measurements where Dirac features do not shift despite increasing pump fluence or time delay, ascribed to a phase transition nucleated in local domains \cite{MogiARX2025}, as well as previous structural probe studies noting domain formation \cite{WangNPJQM2021} after photoexcitation. Finally, this scenario is compatible with theoretical predictions \cite{MocattiJPCL2023}
finding a fluence regime where $Immm$ and $Pnma$ are both dynamically stable.

\indent The observed high-frequency enhancement dynamics are only well-resolved for the two highest measured fluences, again consistent with a threshold to the 800~nm-pumped phase dynamics in SnSe \cite{DringoliPRL2024,HuangPRX2022}. This dynamic plasmonic-type enhancement above all phonon energies is a known response of phase segregation, where local domains create depolarization fields that have a Lorentzian resonance at a renormalized collective plasma frequency \cite{KuzelJPD2014,JoyceSST2016}. These effects could be described by an effective medium theory such as the Bergman model \cite{BergmanPR1978}, but such theories consider systems containing metallic domains in an insulating bulk. The photoexcitation employed here both affects the proposed photoexcited domains and brings metallic character to the $Pnma$ bulk, so such models cannot be directly applied, but the presence of low frequency conductivity suppression and mixed phonon responses makes an effective medium treatment possible in principle.

\begin{figure}[b!]
    \centering
    \includegraphics[width=0.85\columnwidth,trim={0.2cm 0.5cm 1cm 2.5cm},clip]{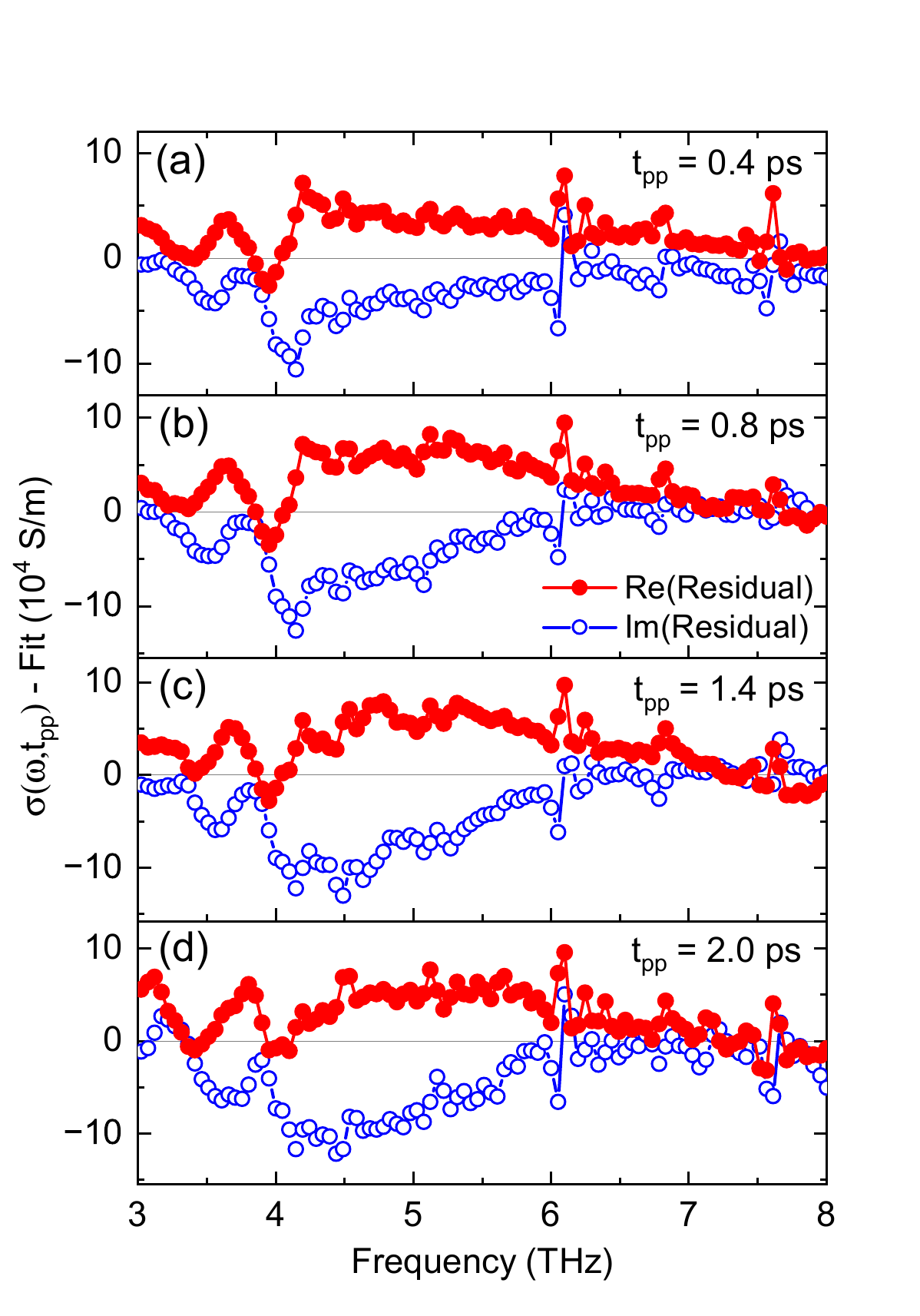}
    \caption{(a-d) Fit residuals for small pump-probe delays at 3.1~mJ/cm$^2$ pumping strengths. The real and imaginary conductivity shape is strongly reminiscent of Dirac interband conductivity \cite{BasovNP2008,NeubauerPRB2016,HorngPRB2011}, but full dynamics are difficult to capture as the high-frequency Lorentzian feature redshifts into this region at longer pump-probe delays.}
    \label{3.1Residuals}
\end{figure}

\indent The observed phonon feature dynamics with increasing pump-probe delay and excitation density supports a photoinduced phase transition pathway from the $Pnma$ phase towards the higher-symmetry $Immm$ lattice orientation. Both a restoration of phonon degeneracy and a reduction in phonon anharmonicity can cause linewidth narrowing effects. The blueshift of the B$^2_{1u}$-identified feature with increasing fluence aligns with $Immm$ SnSe showing a more degenerate optical phonon mode network at higher frequencies compared to $Pnma$ when photoexciting above a critical charge carrier density \cite{MocattiJPCL2023}. This observation of higher-energy phonon features is counter to that expected for a transition towards $Cmcm$, where a redshift of the phonon network has been observed \cite{BansalPRB2016}. This effect is also reflected in the DFT simulations shown in Fig.\ref{SFig5}, but the interplay between photoexcited and background conductivities makes connecting the observed feature blueshift with real phonon mode changes difficult. The most dramatic center frequency changes occur in the same time region (2 - 5~ps) as the narrowing and redshift dynamics of the high frequency Lorentzian response shown in Fig. \ref{LorentzFitParams}(b), and therefore links the phonon dynamics to the relaxation of excited charge in the proposed nucleated regions. Given the strong electron-phonon coupling in SnSe \cite{CarusoPRB2019}, this implies that the photoinduced phonon changes are initiated by the same electronic excitations responsible for the plasmonic response related to phase heterogeneity. The dynamic lattice response therefore stabilizes as carriers energetically relax within their local domains.\\
\indent Finally, inspecting the residuals of the 3.1~mJ/cm$^2$ fits at early pump-probe delays in Figure \ref{3.1Residuals} shows a positive step in $\sigma_1(\omega)$ and a negative peaked $\sigma_2(\omega)$, reminiscent of interband conductivity in Dirac band materials like graphene \cite{BasovNP2008,NeubauerPRB2016,HorngPRB2011,FalkovskyPRB2007}. An example of Dirac interband conductivity at different temperatures is shown in Supplemental Figure 5. This observation aligns with the presence of in-gap Dirac states recently reported under photoexcitation conditions similar to those presented here \cite{MogiARX2025}. Additionally, the reduced SnSe thickness in these transmission measurements compared to previous reflection measurements increases sensitivity to these surface states. Despite this, the captured features here are mixed with the adjacent phonon and plasmonic contributions, making quantitative interpretation difficult. Its appearance at 3.1~mJ/cm$^2$, where thresholds in other features also appear, gives further support to a thresholded process occurring near 3~mJ/cm$^2$. This threshold fluence value is lower than in our previous work on bulk single crystal SnSe (6.6~mJ/cm$^2$) where carrier diffusion into the bulk can occur, but better matches the predicted carrier density threshold for photoinduced phase transition in SnSe \cite{HuangPRX2022}. The observation that this step in conductivity appears at the strongest-coupled optical phonon energy may indicate the involvement of phonons via an indirect interband transition between Dirac cone states. 

\section{Conclusions}
\label{sec:Conclusions}

\noindent The ultrabroadband THz photoconductivity spectra presented here reveal charge carrier localization through dynamic interruption of long range conduction occurring on a several ps time scale. During this time the lattice migrates to a non-equilibrium response, with phonon feature narrowing and shifting on similar time scales, supporting local domains transitioning towards $Immm$ SnSe. Observation of a new phonon mode is consistent with DFT calculations of a transient mixed phase of $Pnma$, $Immm$ and at long delays following carrier recombination $Fm\bar{3}m$ however confirmation will require a spatially resolved technique. A plasmonic peak at high frequencies also redshifts on the same time scale, showing carrier relaxation within such a heterogeneous phase landscape occurring on longer picosecond time scales. The threshold for these signatures of phase heterogeneity appear at 3.1~mJ/cm$^2$, lower than previous measurements on bulk crystals. The concomitance of different subsystem dynamics shows the critical interplay between lattice structure and electronic perturbations in photoexcited SnSe. Further experiments using THz-Scanning Tunneling Microscopy \cite{CockerNAT2021} or other spatially-resolved probes could allow for confirmation of the nucleated nature of this photoexcited phase transition and further elucidate the dynamics of each phase independently.

\section{Acknowledgments}
This work was supported through funding by NSERC and Mitacs programs. The Northwestern University personnel and research work was supported by the Department of Energy, Office of Science, Basic Energy Sciences under Grant No. DE-SC0024256 (design and synthesis of thermoelectric materials). 

We acknowledge the CINECA award under the ISCRA initiative (projects IscrC\_Uf--DynFP and IscrB\_EPhoCS), for the availability of high performance computing resources and support. We acknowledge EUROHPC (project 465000468) for the availability of high performance computing resources and support.

Funded by the European Union (ERC, DELIGHT, 101052708). Views and opinions expressed are however those of the authors only and do not necessarily reflect those of the European Union or the European Research Council. Neither the European Union nor the granting authority can be held responsible for them.

\bibliography{SnSe_Mar23_Abbrv}

\end{document}

\title{Photoinduced phase heterogeneity and charge localization in SnSe: \\ Supplemental Information}
\author{Benjamin J. Dringoli}
    \affiliation{Department of Physics, McGill University, Montreal, QC, Canada H3A2T8}

\author{Stefano Mocatti}
    \affiliation{Department of Physics, University of Trento, Via Sommarive 14, 38123 Povo, Italy}

\author{Giovanni Marini}
    \affiliation{Department of Physics, University of Trento, Via Sommarive 14, 38123 Povo, Italy}

\author{Zhongzhen Luo}
    \affiliation{Department of Chemistry, Northwestern University, Evanston, IL 60208, United States}
    \affiliation{Key Laboratory of Eco-Materials Advanced Technology, College of Materials Science and Engineering, Fuzhou University, Fuzhou 350108, P. R. China}

\author{Matteo Calandra}
    \affiliation{Department of Physics, University of Trento, Via Sommarive 14, 38123 Povo, Italy}

\author{Mercouri G. Kanatzidis}
    \affiliation{Department of Chemistry, Northwestern University, Evanston, IL 60208, United States}

\author{David G. Cooke}
    \email[Correspondence email address: ] {david.cooke2@mcgill.ca}
    \affiliation{Department of Physics, McGill University, Montreal, QC, Canada H3A2T8}

%\date{\today} % Leave empty to omit a date

\maketitle

\section{Differential THz conductivity extraction procedure}

\begin{figure}[h!]
    \centering
    \includegraphics[width=0.7\columnwidth,trim={0cm 1.5cm 2cm 6cm},clip]{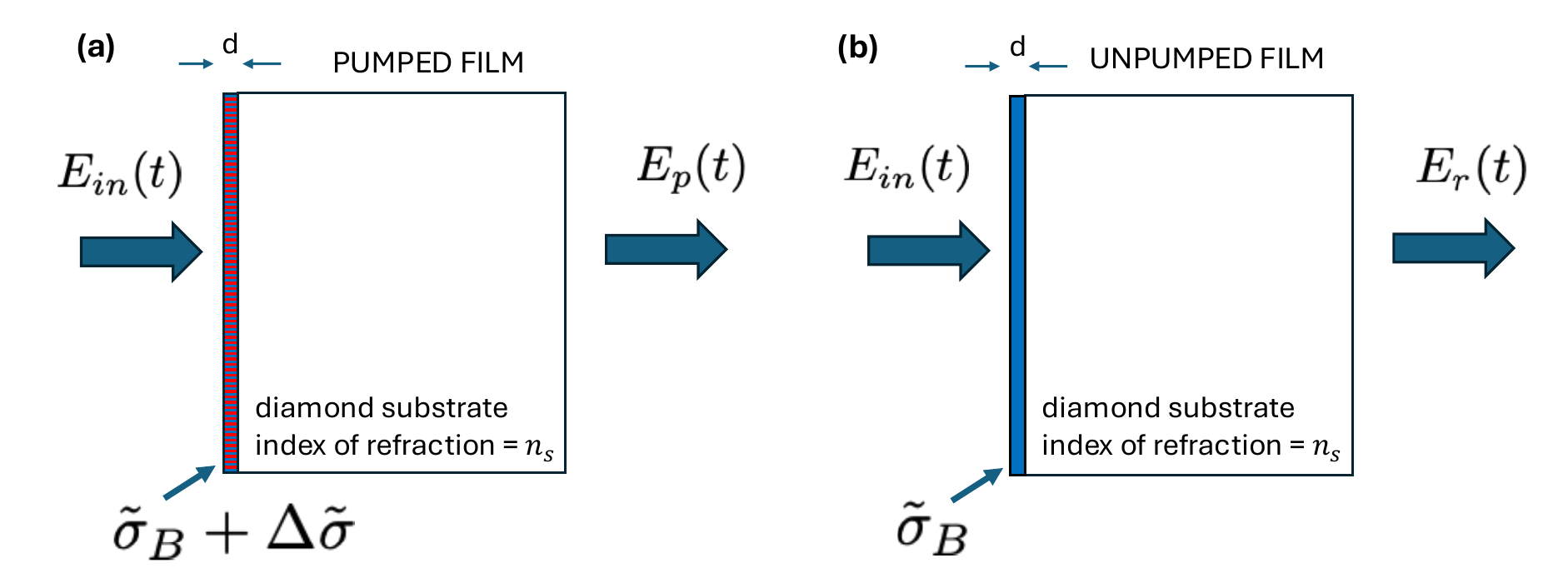}
    \caption{(a) TRTS experiment with pump active, contributing $\Delta \tilde{\sigma}$} in addition to the (b) background $\tilde{\sigma}_B$ present in the unpumped case.
    \label{SFig1}
\end{figure}
In the THz transmission experiment, the THz electric fields are time sampled in both pumped $E_{p}(t)$ and unpumped $E_r(t)$ cases, shown in Fig.~\ref{SFig1} (a) and (b), respectively. The goal of the experiment is to measure the pump-induced differential complex conductivity $\Delta\tilde{\sigma}$ of the thin film material with thickness $d$ on a substrate with index of refraction $n_s$. In many cases, the background conductivity due to background charge carriers or optically active phonons, $\tilde{\sigma}_B$, can be ignored as they are too weak to detect given the thin film thickness. However, in strongly polar phonon systems, this is not the case.

As shown in the main text, to convert the measured transmission into optical conductivity we use a Tinkham thin film approximation. In the presence of significant background conductivity $\tilde{\sigma}_B$, this leads to a differential transmission of the form

\begin{equation}
    \tilde{t}(\omega) = \frac{E_p / E_{in}}{E_r / E_{in}} = \frac{2 / (1+n_s+Z_0 \tilde{\sigma} d)}{2 / (1+n_s+Z_0 \tilde{\sigma}_B d)} = \frac{1+n_s+Z_0 \tilde{\sigma}_B d}{1+n_s+Z_0 \tilde{\sigma} d}.
\end{equation}
where $\tilde{\sigma} = \tilde{\sigma_B} + \Delta\tilde{\sigma}$. The extracted conductivity now strongly depends on the details of $\tilde{\sigma}_B$. In principle this background contribution could be measured with sufficient accuracy to extract the differential conductivity without artifacts, however in practice the sharp phonon features and dynamic evolution of their center frequency and lineshape in the photoexcited state makes such a sensitive inversion impossible. Therefore, we choose to analyze the changing background influence directly by setting $\tilde{\sigma}_B = 0$, effectively absorbing its contributions into the differential conductivity and leading to Equation 2 in the main text.

\begin{figure}[t!]
    \centering
    \includegraphics[width=0.8\columnwidth]{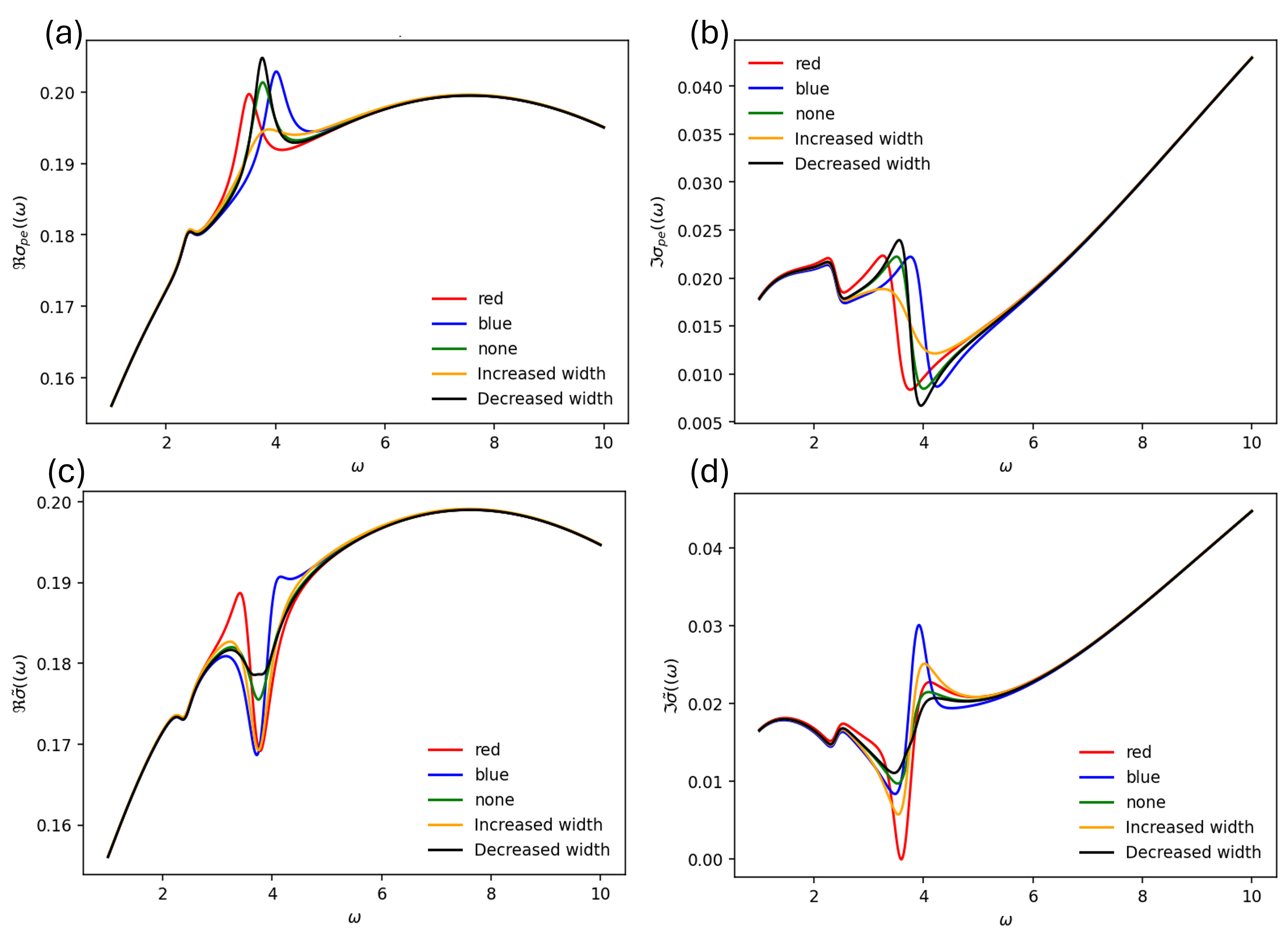}
    \caption{Simulated (a) real and (b) imaginary parts of the optical conductivity including a Drude-Smith intra-band component and $B^1_u$ and $B^2_u$ Lorentzian phonon resonances. (c) and (d) show the corresponding real and imaginary conductivity obtained by ignoring the background phonon conductivity. Lines correspond to a hypothetical redshift (red), blueshift (blue), and no shift (green) in resonance frequency as well as an increase (yellow) and decrease (black) in $B^2_u$ linewidth.}
    \label{SFig2}
\end{figure}

The main effect of ignoring the background conductivity is the appearance of negative Lorentzian contributions centered on the background infrared-active phonons in the extracted photoconductivity. Care must be taken in the physical interpretation of the Lorentzian fits to recorded $\tilde{\sigma}(\omega) = \tilde{\sigma}_B(\omega)+\Delta \tilde{\sigma}(\omega)$ features however, as a blue shift of the Lorenztian fits need not correspond to a blue shift of the phonon frequency itself. Indeed, we have modeled the effect of a pump-induced red- and blue-shift, and a pump-induced increase and decrease in the phonon linewidth of the $B^2_{1u}$ phonon in Fig.~\ref{SFig2}. Fig.~\ref{SFig2}(a) and (b) show the simulated real- and imaginary parts of the complex conductivity on the background of a broad Drude-Smith conductivity. Fig.~\ref{SFig2}(c) and (d) show the corresponding real and imaginary differential conductivity resulting from Fig.~\ref{SFig2}(a) and (b) using the Tinkham assumption and ignoring the background conductivity. One can see that both the pump-induced red- and blue-shift of the phonon leads to a narrowing of the extracted Lorentzian dip in the conductivity, with the appearance of an asymmetric lineshape. Moreover, and somewhat counter-intuitively, an increased width of the phonon can produce a decrease in the Lorentzian dip linewidth and vice versa. We are, however, unable to reproduce the frequency shifts of the Lorentzian dips to the degree observed in the experimental data with any pump-induced change in the phonon linewidth or center frequency shift assuming a symmetric Lorentzian response. We conjecture a more complicated line shape may be photoinduced, e.g. a Fano lineshape, however more detailed temperature measurements of the background conductivity than is possible here are required to verify.

\newpage

\section{Additional early-time TRTS measurements}
\begin{figure}[h!]
    \centering
    \includegraphics[width=0.45\columnwidth,trim={0cm 0.5cm 0.5cm 1.1cm},clip]{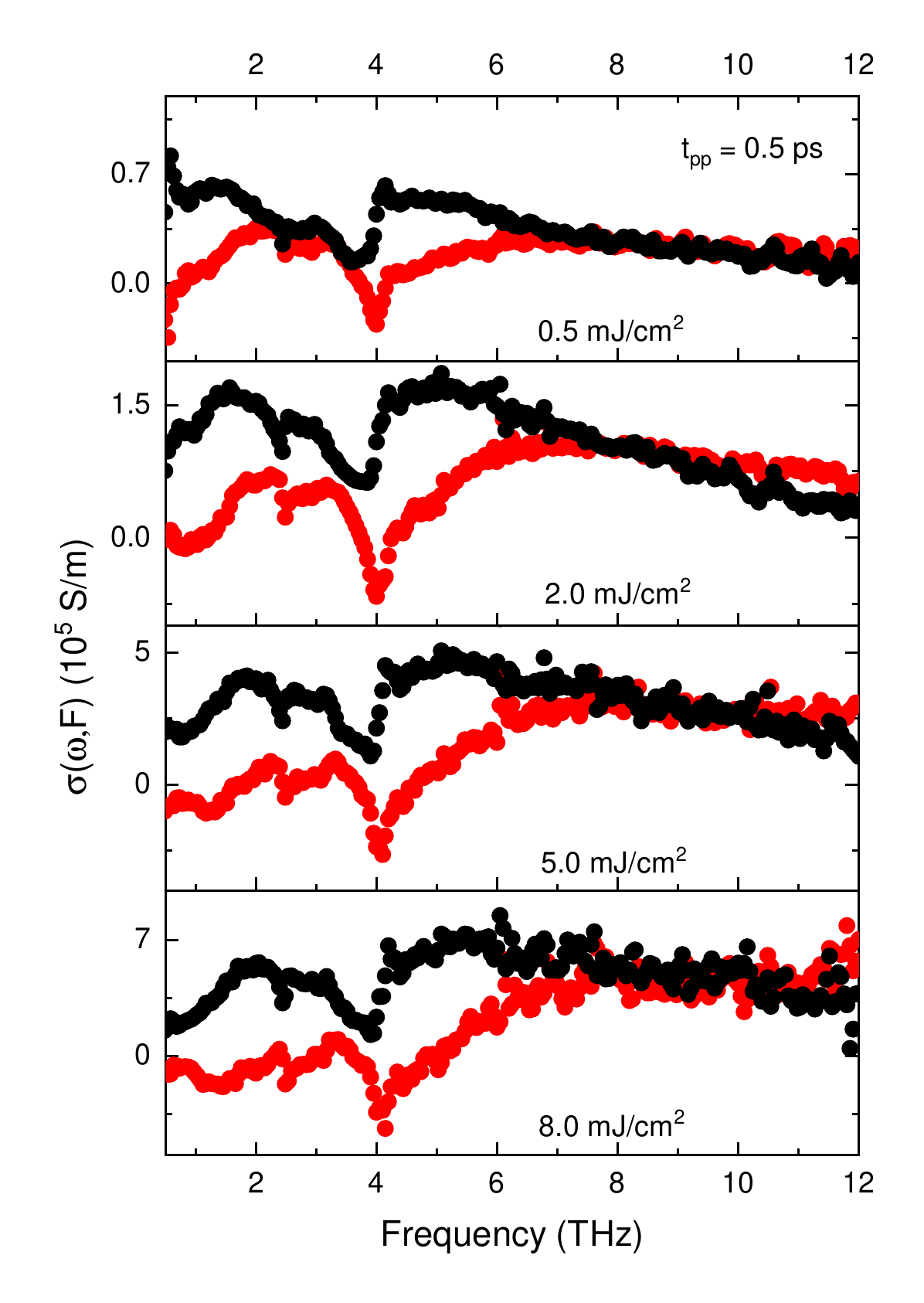}
    \includegraphics[width=0.42\columnwidth,trim={0cm 0.5cm 0.5cm 1cm},clip]{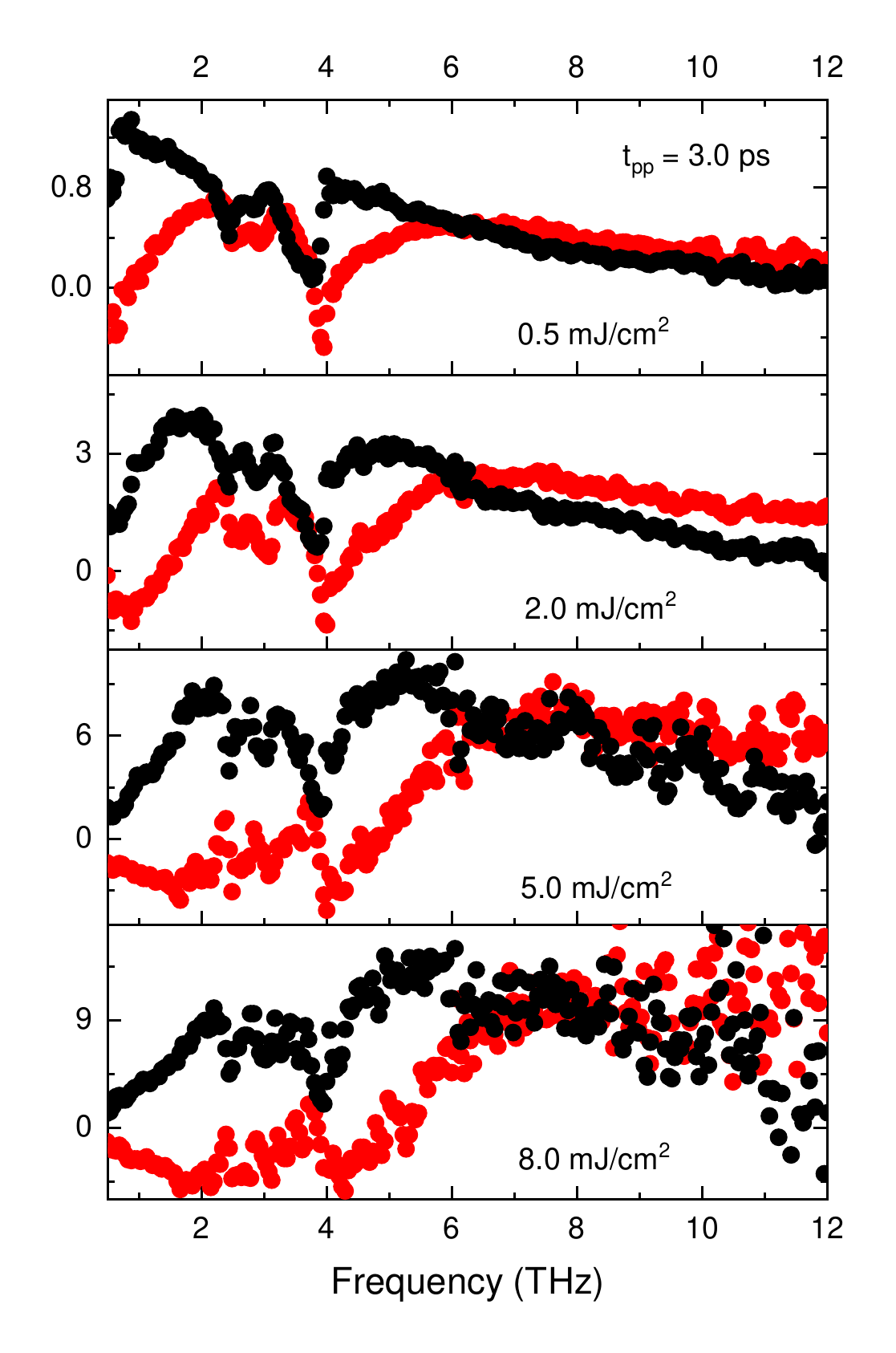}
    \caption{Fluence-dependent complex conductivity $\sigma(\omega,F,t_{pp})$, $\sigma_1$ black and $\sigma_2$ red, for pump-probe delays of 0.5~ps (left) and 3.0~ps (right). Suppression of the zero-frequency conductivity is seen with increasing fluence at both pump probe delays. The 3.0~ps data also demonstrates the B$^2_{1u}$ response narrowing with increasing fluence at slightly later pump delays. Such narrowing is not yet observed at $t_{pp}=0.5$~ps.}
    \label{SFig3}
\end{figure}
Additional time-resolved THz spectroscopy measurements were performed for early pump-probe delay times to confirm repeatability and sample uniformity, shown in Supplemental Fig.~\ref{SFig3}. 
The noted low-frequency optical conductivity suppression recovered in the TRTS data was reproduced in separate fluence-dependent measurements on the same SnSe sample. These plots also reproduce the same phonon linewidth narrowing with increasing fluence and pump-probe delay, and show that it is only realized after a few ps post-excitation, as the features at 0.5~ps are similarly broad for all excitation fluences.
\begin{figure}[h!]
    \centering
    \includegraphics[width=0.8\columnwidth,trim={0cm 0cm 0.5cm 1.5cm},clip]{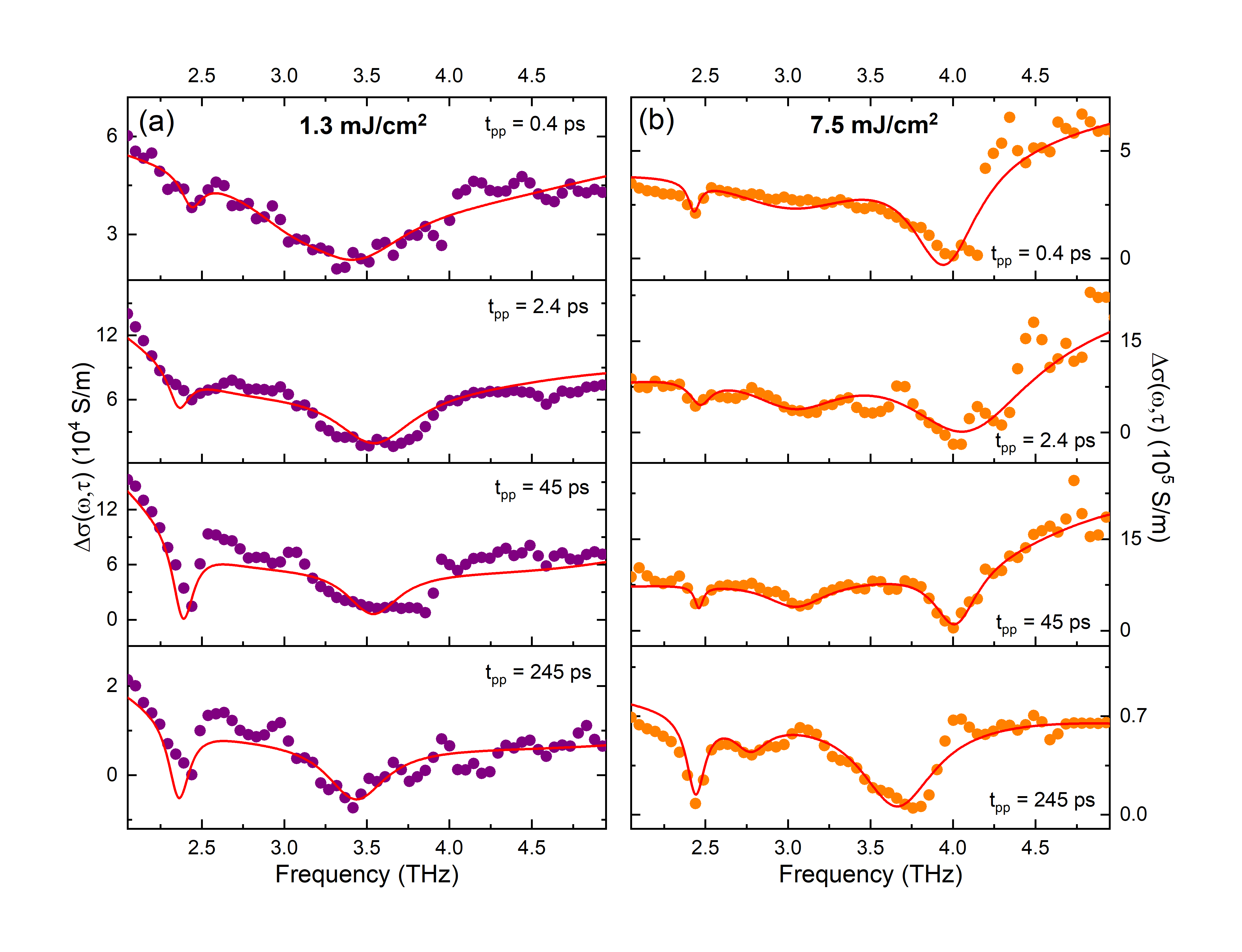}
    \caption{Selected fits and extracted data showing the poor linewidth fitting results, particularly for the (a) 1.3 and (b) 7.5 mJ/cm$^2$ pump fluence experiments.}
    \label{SFig4}
\end{figure}
%\begin{figure}[h!]
%    \centering
    %\includegraphics[width=0.5\columnwidth,trim={0cm 0.5cm 0.5cm 1.5cm},clip]{Phonon2LinewidthFail.pdf}
    %\caption{Dynamics of the $B^2_{1u}$ Lorentzian damping parameter $\gamma_{2}$ across different fluences. Fits are unreliable for higher fluences due to extensive broadening, particularly 7.5~mJ/cm$^2$.}
    %\label{LinewidthFail}
%\end{figure}
The phonon fits using Lorentzian functions are unable to account for the asymmetric phonon lineshapes, particularly appearing at higher pump fluences. This is shown in Fig.~\ref{SFig4}, where the linewidths are seen to be poorly described (e.g. $t_{pp}=2.4$~ps). The Lorentzian fits are therefore only used to quantitatively describe the center frequency dynamics.
\newpage

\newpage

\section{Residual Feature}

\indent Fig.~6 of the main manuscript shows the residuals of the complex conductivity fits for F = 3.1~mJ/cm$^2$ between $4-7$~THz and $0 < t_{pp} < 2$ ps. Both the real and imaginary conductivity of this residual feature match what is expected from allowed interband transitions in Dirac-like systems, taking the form

\begin{equation}
    \tilde{\sigma}(\omega)=C*\frac{\omega+i \Gamma}{\pi} \bigg[ \textup{ln} \bigg| \frac{\hbar \omega+ i \hbar \Gamma + 2 \Lambda}{\hbar \omega+ i \hbar \Gamma - 2 |\mu|} \bigg| +i \ \textup{arctan} \bigg( \frac{\hbar \omega+ i \hbar \Gamma - 2 |\mu|}{\hbar \Gamma} \bigg) \bigg].
\end{equation}

\noindent Here $\Lambda$ is a high-energy cutoff, $\mu$ is the chemical potential, and $\Gamma$ is the scattering rate \cite{FalkovskyPRB2007}. \\
\indent The measured deviation is short-lived and weak in comparison to the other features present in this system, making deep analysis difficult. The onset of conductivity in this system is typically determined by inter-band transitions across the Dirac cone, and so therefore would typically be sensitive to the quasi-Fermi level of the system. We, however, observe the onset of the residual feature is approximately constant at $\sim4$~THz = 16.5~meV as charge carriers in the system relax. We conjecture that this component of the conductivity could be due to phonon assisted interband transitions across the newly formed topological Dirac-like bands.

\begin{figure}[h!]
    \centering
    \includegraphics[width=0.75\columnwidth,trim={0cm 1cm 0cm 0.2cm},clip]{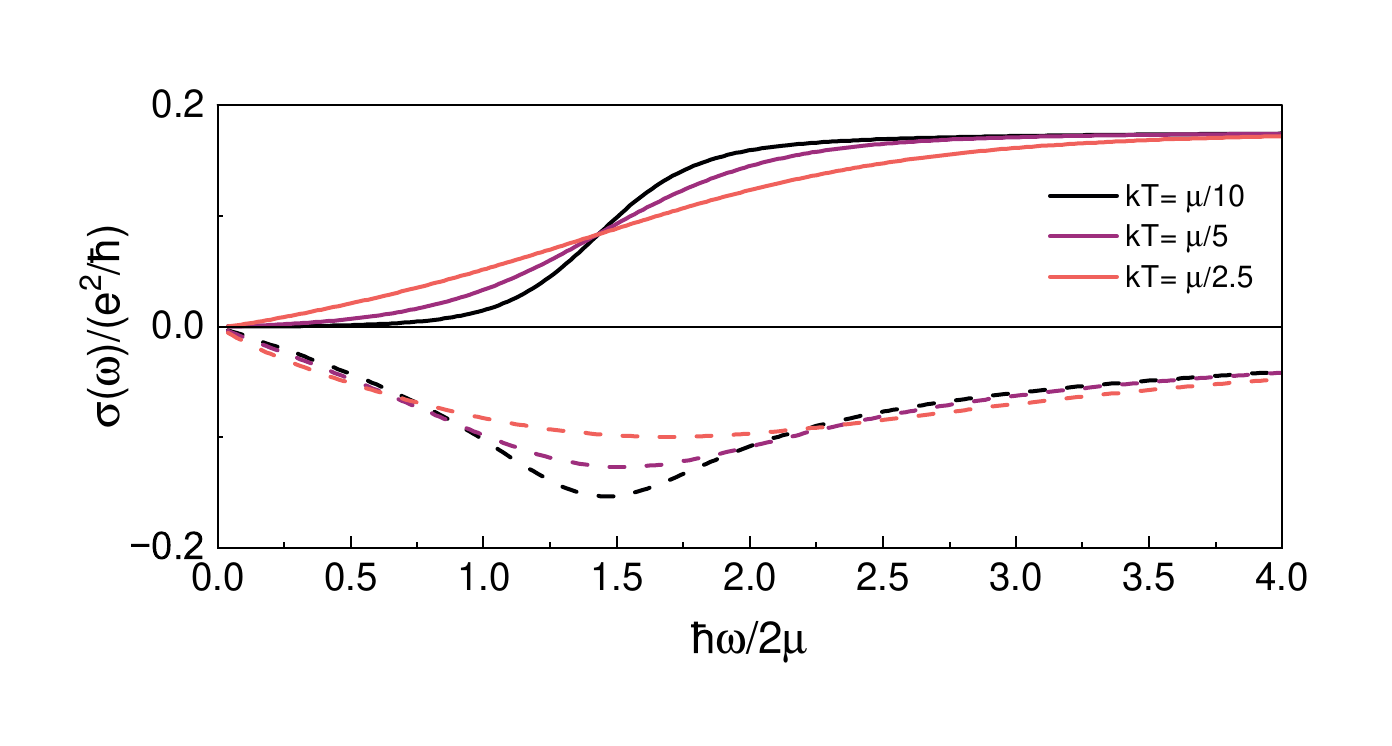}
    
    \caption{Dirac interband conductivity as a function of temperature. The trend to wider, more spread-out features with increasing temperatures qualitatively follows the dynamics seen in the $t_{pp} < 2$ ps residual data show in Figure 6 of the main text.}
    \label{SFig6}
\end{figure}

\newpage

\section{Determination of the sample thickness}
\noindent The thickness of the exfoliated SnSe sample was measured by atomic force microscopy, shown in Supplemental Figure \ref{AFM}. Multiple line cuts show an average film thickness of approximately 500~nm. This measurement was performed on the diamond substrate used for THz spectroscopy measurements after exfoliation and transfer from the bulk crystal, and so is identical to what is measured in the THz experiments.
\begin{figure}[h!]
    \centering
    \includegraphics[width=0.85\columnwidth,trim={5.5cm 5cm 4cm 5cm},clip]{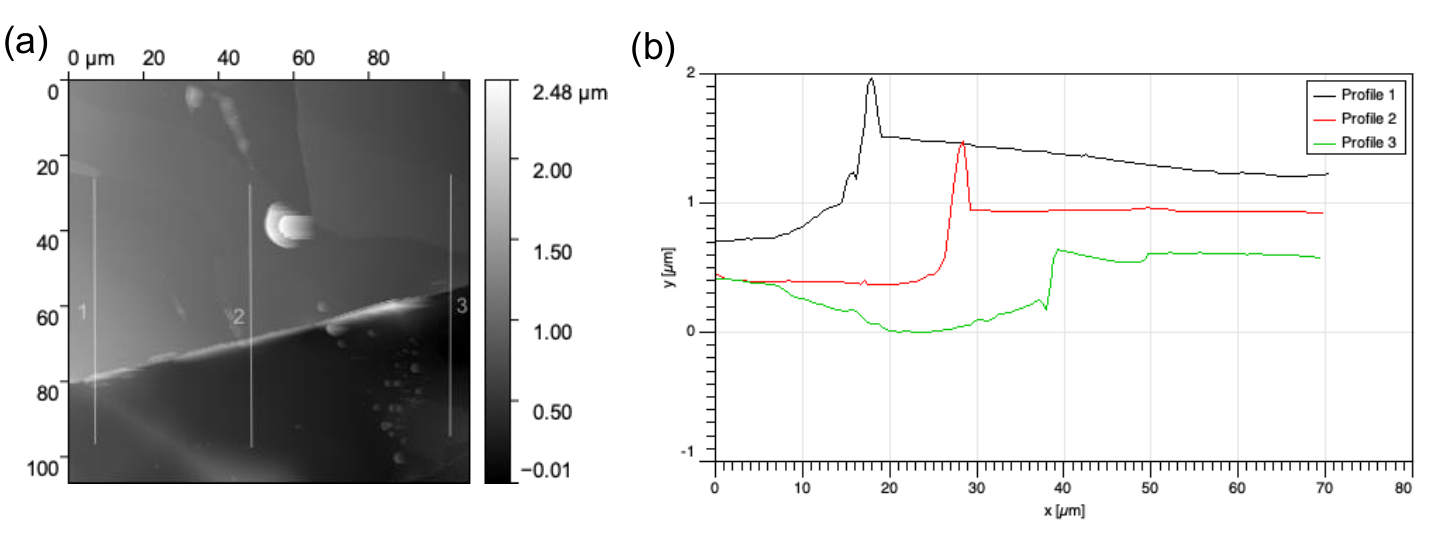}
    \caption{(a) Atomic force microscopy image with line cuts showing exfoliated SnSe on the diamond substrate. (b) Plotted height variation along noted line cuts. The average SnSe thickness is approximately 500~nm.}
    \label{AFM}
\end{figure}

\bibliography{bibliography}